\documentclass{SciPost}

\hypersetup{
    colorlinks,
    linkcolor={red!50!black},
    citecolor={blue!50!black},
    urlcolor={blue!80!black}
}

\usepackage[bitstream-charter]{mathdesign}
\DeclareSymbolFont{usualmathcal}{OMS}{cmsy}{m}{n}
\DeclareSymbolFontAlphabet{\mathcal}{usualmathcal}

\fancypagestyle{SPstyle}{
\fancyhf{}
\lhead{\colorbox{scipostblue}{\bf \color{white} ~SciPost Physics }}
\rhead{{\bf \color{scipostdeepblue} ~Submission }}

\fancyfoot[C]{\textbf{\thepage}}
}

\usepackage{graphicx,amsfonts,amsmath,hyperref,multirow,array}
\usepackage[utf8]{inputenc}

\newif\ifhyper
\hypertrue
\ifhyper
\hypersetup{
  citecolor = {green},
  colorlinks = {true}, 
  urlcolor = {blue} 
}
\fi

\newlength{\ldag}
\begin{document}

\pagestyle{SPstyle}

\begin{center}{\Large \textbf{\color{scipostdeepblue}{
Large scale behavior in the Kuramoto-Sivashinsky equation: The Schwinger-Dyson route\\
}}}\end{center}

\begin{center}\textbf{
O. Coquand\textsuperscript{1$\star$}
}\end{center}

\begin{center}
{\bf 1} Laboratoire de Modélisation Pluridisciplinaire et Simulations, Université de
Perpignan Via Domitia, 52 avenue Paul Alduy, F-66860 Perpignan, France
\\[\baselineskip]
$\star$ \href{mailto:email1}{\small oliver.coquand@univ-perp.fr}
\end{center}

\section*{\color{scipostdeepblue}{Abstract}}
\textbf{\boldmath{%
The present paper is a study of the large scale properties of the Kuramoto-Sivashinsky equation.
By using a Schwinger-Dyson framework, we aim to provide a proof that the only solutions that can sustain a stable
scaling in the infrared limit have a negative effective viscosity (in the Kuramoto-Sivashinsky sense) with a minimal
set of hypotheses, thereby showing that this constitutes a general property of the wave equation that does not depend
on a specific set of truncations of a renormalisation group flow, or limitations of a given numerical scheme for example.
}}

\vspace{\baselineskip}

\noindent\textcolor{white!90!black}{%
\fbox{\parbox{0.975\linewidth}{%
\textcolor{white!40!black}{\begin{tabular}{lr}%
  \begin{minipage}{0.6\textwidth}%
    {\small Copyright attribution to authors. \newline
    This work is a submission to SciPost Physics. \newline
    License information to appear upon publication. \newline
    Publication information to appear upon publication.}
  \end{minipage} & \begin{minipage}{0.4\textwidth}
    {\small Received Date \newline Accepted Date \newline Published Date}%
  \end{minipage}
\end{tabular}}
}}
}


\vspace{10pt}
\noindent\rule{\textwidth}{1pt}
\tableofcontents
\noindent\rule{\textwidth}{1pt}
\vspace{10pt}



\section{Introduction}

	One of the most studied partial differential equation in physics is certainly the Navier-Stokes equation.
	Among its many properties, a striking one is the possibility for a system obeying this equation to spontaneously develop a form of stochastic behavior,
	even if stochasticity is absent both from the initial condition and the evolution equation; such solutions correspond to turbulent flows.
	The Navier-Stokes equation is however not the only one in that case.
	A number of other equations presenting the property of so-called \textit{spontaneous stochasticity} have been introduced over the years,
	often with the aim of presenting models deemed simpler than the Navier-Stokes turbulence one.

	One example of such an equation is the Kuramoto-Sivashinsky (KS) equation.
	It was initially introduced by Sivashinsky to describe the perturbation of a flame front near the laminar regime \cite{Sivashinsky77} from the
	fundamental equations of flame front propagation \cite{Barenblatt11}.
	In the appropriate asymptotic regime, it can be written as follows:
	\begin{equation}
	\label{eqKSKPZ}
		\partial_t h + \nu \partial^2 h + \zeta \partial^4 h + \frac{1}{2}\big(\partial_i h\big)^2 = f(x)
	\end{equation}
	where $i$ as all latin indices in the following refers to spatial directions only, $f(x)$ is a fluctuating term that we do not need to specify so far
	and $x$ is a four-vector composed of time and space (spatial vectors are in bold font).
	This form (\ref{eqKSKPZ}) will be referred to as the KS-KPZ form of the equation in the following, because in the appropriate asymptotic limit it takes
	a form very close to that of the Kardar-Parisi-Zhang (KPZ) equation (see discussion below).
	Eq.~(\ref{eqKSKPZ}) is the most general form of the flame front fluctuation equation; other asymptotic forms where one of the constants on the left-hand side
	are equal to zero are also discussed in the original paper.

	In a later study \cite{Sivashinsky80}, Sivashinsky showed that the motion of a liquid film down a vertical plane obeys a very similar equation:
	\begin{equation}
	\label{eqKSB}
		\partial_t u_j + \mu \,u_j\partial_j u_i + \nu \partial^2 u_i + \zeta \partial^4 u_i = 0
	\end{equation}
	where, in the multidimensional version of the equation, the velocity field obeys the further constraint $\varepsilon_{ijk}\partial_ju_k=0$ ($\varepsilon$ being
	the Levi-Civita connection).
	In the limit where $\zeta\rightarrow0$, and up to a sign in front of $\nu$,
	this equation becomes the (multidimensional) Burgers equation; hence we will call it the KS-B form of the KS equation.
	It is interesting to note that, if we forget about the noise term, the KS-B equation can be deduced from the KS-KPZ form by the transformation
	$u_i=\partial_ih$.

	The work of Kuramoto, done in a similar time window \cite{Kuramoto78}, is concerned with chemical turbulence, and more precisely, the development of chaos, or stable
	limit cycles, in systems of chemical substances interacting together.
	In order to study such systems, Kuramoto first introduced a phase field $\phi(x)$, that represents the approximate value of the time period of the concentration
	field in phase space.
	He then showed on very general grounds that, in the limit of large distances, the phase field can be shown to obey the following equation:
	\begin{equation}
	\label{eqKu}
		\partial_t\phi + \nu \partial^2\phi + \zeta \partial^4\phi+\mu\big(\partial_i\phi\big)^2=0
	\end{equation}
	which is nothing but the (KS-KPZ) equation without forcing; this regime is thus called the \textit{phase turbulence} regime.
	Obviously, taking one spatial derivative, one gets back the (KS-B) form of the equation.

	In his first paper \cite{Kuramoto78}, Kuramoto already noted that the Laplace operator term coming with an unusual sign (it would correspond to a negative viscosity in
	a Burgers or Navier-Stokes fluid) is at the origin of the transformation of the solutions from travelling waves if $\nu<0$ to chaotic solutions if
	$\nu>0$.
	Let us already note at this stage that this happens while in the Kuramoto's version of the equation, there is no stochastic forcing term in the right-hand side
	of the equation.

	Pushing his study further, Kuramoto also showed that another asymptotic study could be performed on the equations of chemical mixing, which then take the form:
	\begin{equation}
	\label{eqKCGLE}
		\partial_t w - w +(1+i\,c_1)\partial^2w + (1 + ic_2)\big|w\big|^2w = 0
	\end{equation}
	which has the form of a complex Ginzburg-Landau equation that is also used to describe driven-dissipative bosonic condensates \cite{Vercesi24}, thereby exemplifying
	the richness of the physics described by the equations in the KS class.

	In order to map this equation to the previous formulation, one should apply the transformation $\nu=-(1+c_1c_2)$, {$\zeta=c_1^2(1+c_2^2)/2>0$}.
	For large positive values of $\nu$, a new type of instability can then develop in the equation, that Kuramoto called \textit{amplitude turbulence}.

	Since then, a large number of studies have tried to unveil the properties of the KS equation, particularly regarding the emergence of chaotic behavior even
	in the Kuramoto version of the equations, that is, without the presence of any stochastic forcing.
	The complete analysis of the possible solutions depending on the value of the parameters, and the associated bifurcations lead a particularly rich and complex
	state diagram (see Figs~1-3 of \cite{Hyman86}).
	Moreover, the kind of chaos emerging from the solutions of the KS equation appears to be different from that which is encountered in Lorenz model for
	example \cite{Datta26}: the Lorenz model belongs to the class of ordinary differential equation chaos, while the KS model belongs to the class of partial
	differential equation chaos, being truly chaotic both in time and in space.

	Among the well established properties of the KS equation (in both formulations) is the fact, that can be derived by linear analysis, that the unusual Laplace operator
	term leads to an instability of the solutions and their progressive degeneracy into chaos, while the non-linear term tends to regularise the solution \cite{Fujisaka77}.
	Then, a seminal paper in the domain is the first attempt by Yakhot \cite{Yakhot81} to derive the scaling properties of the statistical moments of the
	stochastic solution by use of renormalisation group tools.
	Unfortunately, his renormalisation group procedure turns out to be very cumbersome, introducing an explicit ultraviolet (UV) scale separating the fast from the slow
	modes, and rendering thus doubtful the ability of the resulting procedure to describe correctly the infrared (IR), large distance, properties of the solutions.
	Moreover, the main result of this study was the fact that in 1+1 dimensions, $\nu$ changes sign upon renormalisation, but this does not hold in higher dimensions.
	This has confined most of the successive studies of the subject to the 1+1 dimensional case while 2+1 and 3+1 dimensions, at least, are certainly relevant to the
	initial problems posed by Sivashinsky and Kuramoto.

	In an attempt to compare KS, and particularly the KS-B version to Burgers turbulence, Pomeau and his collaborators \cite{Pomeau84} provided an extensive study
	of the scaling of the power spectrum of the solutions to the KS equation, and their stability with respect to the addition of a small perturbation in the equation.
	They have shown that the spectrum has three distinct regimes: (i) a flat regime in the infrared, (ii) a peak at the instability length followed by a robust
	$k^{-4}$ decay and (iii) a final exponential decay.

	A more mathematical study by Nicolaenko and his collaborators \cite{Nicolaenko85} proved that the KS equation in its Kuramoto formulation, that is with a right-hand
	side equal to zero, can be related by a change of variable to an equivalent equation with a time-dependent stochastic noise term instead (this is a weaker
	form than the Sivashinsky equation (\ref{eqKSKPZ}) where the noise is space-time dependent).
	This is significant inasmuch as the question of the extent to which the stochastic forcing plays a role in determining the chaotic properties of the solution
	to the KS equation is a still unanswered question today.

	Then, Zaleski made a detailed numerical study of the scaling properties of the solutions in the IR regime \cite{Zaleski89}.
	In particular, he described the behavior of the dynamical critical exponent $z$ that describes the relative scaling of the time in units of length.
	He showed that $z$ first stabilises to a value $z=2$, corresponding to the Edwards-Wilkinson (EW) model, before crossing over to finally reach the
	value $z=3/2$ corresponding to the 1+1 dimensional KPZ model in the far infrared.
	Another byproduct of this study is the observed behaviour $\nu(\Lambda)\sim\Lambda^{z-2}$, where $\Lambda$ is an energy scale. This confirms that
	(i) the viscosity does indeed increase in the infrared regime, (ii) this growth can be slowed down if the system adopts the EW scaling for a wide range of scales.
	This is to be put in contrast to the predictions of Yakhot \cite{Yakhot81} that the viscosity does not grow in dimensions bigger than one.
	Among the last conclusions of this study is the fact that "the Kuramoto equation behaves at large scales as a Burgers equation with a white noise forcing"
	\cite{Zaleski89}, which means that in the IR, (i) $\zeta$ plays a subdominant role, (ii) the viscosity term changes sign ($\nu$ becomes negative at large scales
	if it was positive at short scales).
	These results were later confirmed by an updated numerical study \cite{Sneppen92}, and a further one where universality with respect to the addition of higher
	powers of the Laplace operator up to $\partial^8$ has been explicitly checked \cite{Hayot93}.

	The next decade was less productive.
	In \cite{Sakaguchi02}, Sakaguchi identified tracks of Burgers shock-waves in the mean profile of KS-B solutions with a stochastic forcing.
	He then tried to provide a Mori-Zwanzig type of analysis of the scaling behavior, unfortunately because he then broke the convolution product of the
	Mori-Zwanzig kernel, it is very difficult to assess the precision of the obtained results.
	Three years later \cite{Ueno05}, Ueno and his collaborators provided a more detailed renormalisation group (RG) analysis than that provided by Yakhot many years ago.
	But here too the results are not so easy to connect with the previous literature.
	In particular it is claimed in this paper than the forcing noise needs to be scale dependent instead of being just a white noise in order to obtain
	"more reasonable RG results" \cite{Ueno05}.
	Unfortunately, this peculiar procedure makes it, again, difficult to compare in details with the previous results.

	More recently, the evolution of computer capabilities saw the confirmation of the first results being established numerically.
	First, Roy and Pandit \cite{Roy20} performed an extensive study of the KS system with a wide variety of initial condition at short scales.
	They have shown not only that the distribution function of the solution does indeed converge to the KPZ statistics, after a possible crossover in the EW regime,
	but also that the distribution functions at large scales reproduces different types of KPZ distribution functions when the initial conditions are varied,
	which, although not a truly exhaustive proof of convergence to KPZ behavior at large scale , is a very strong result.
	Then, the team of Vercesi \cite{Vercesi24} made a study of the KS equation through its formulation as the phase turbulence in the CGLE equation.
	A major result of this article is the detailed study of the limit $\nu\rightarrow0$, and the numerical proof of the existence of a new scaling regime
	with $z=1$, called the inviscid Burgers (IB) regime.

	Finally, the group of Gosteva, Canet and Wschebor made the last significant contributions to the understanding of the KS equation.
	In a first paper \cite{Gosteva26a}, together with Roy, they showed the ubiquity of the IB scaling as an intermediate regime in the KS-KPZ equation,
	both from a theoretical and numerical points of view.
	Then, in \cite{Gosteva26b}, they made an exhaustive renormalisation group study of the KS in 1+1 dimensions, showing for the first time the existence of all the
	listed scaling regime, including the IB one.
	In particular, they showed that the unusual choice of noise term used by Ueno and collaborators is justified in the KS framework by the necessity to restore
	the time-reversal symmetry at the fixed point.
	They also highlighted some problems with the Wilsonian-type approaches used before when dealing with terms such as the fourth-derivative one that do not appear in
	better studied equations like KPZ, but become singular if not treated appropriately in KS.
	Their last result concerns the deterministic limit in which they found that the length at which the KPZ scaling sets in is sent to infinity.

	What this long review of previous results shows is that despite its apparent simplicity, the behavior of the KS equation turns out to be particularly rich and complex.
	If nowadays, some consensus begins to appear in the literature, the study of this problem has also been characterised by a large confusion, the design
	of cumbersome tools to try to patch for the difficulties of treating the problem as it stands.

	This is the reason why, in this article, we try to provide a view on the problem that is as agnostic as it can be, that is, with a minimal set of assumptions,
	we want to prove that at large scales, the effective viscosity of the KS equation does indeed always change sign.
	In order to do so, we will make only one hypothesis, namely, that \textit{there exists a scaling regime that is stable in the IR limit}.
	This assumption will then be combined with the (exact) Schwinger-Dyson equation corresponding to the KS problem.
	Of course, being so restrictive means that we will not be able to answer all the questions surrounding the KS equation.
	In particular, this way of proceeding hides what can appear at intermediate scales for example.
	Also, in order to establish our Schwinger-Dyson equation, we are going to work directly on the Sivashinsky problem, that is on the forced equation, thereby putting
	aside the question of knowing to what extent the spontaneous stochasticity possess the same characteristics as the one generated by a stochastic forcing.

	The paper is organised as follows.
	First, we present the general setup.
	Then, we begin by presenting the computation procedure in the limiting case of the EW regime, before applying it to the KS case.
	Finally, we conclude

\section{General Setup}

	\subsection{A field theory for the KS problem}

		In order to be able to phrase the problem in a Schwinger-Dyson framework, the first thing we need is to transform the dynamical equation (\ref{eqKSKPZ})
		into a field theory.
		To do that, we follow the Martin-Siggia-Rose-Janssen-De Dominicis (MSRJD) method \cite{Martin73,Janssen76,Dominicis76}.
		At this stage, we need to decide what the properties of the noise are.
		This is done by identifying a precise asymptotics for the equation.
		In our case, we want to enforce that in the limit $\nu<0$, $\zeta\rightarrow0$, $k\rightarrow0$
		(where $k$ is the inverse of the length scale at which the system is probed),
		the system becomes equivalent to a KPZ system, this yields:
		\begin{equation}
			\left<f(x)f(x')\right> = D(x)\,\delta(x-x')
		\end{equation}

		Under such conditions, the MSRJD action becomes:
		\begin{equation}
			\begin{split}
				\mathcal{S}[\varphi] = \int_x\bigg\{\overline{h}(x)\Big[\partial_th(x) + \nu\partial^2h(x) + \zeta\partial^4h(x) 
						      + \frac{1}{2}\big(\partial_ih(x)\big)^2\Big] - D(x)\overline{h}(x)^2\bigg\}
			\end{split}
		\end{equation}
		where $\varphi$ is a generic notation to denote any type of field in the action, and $\int_x$ is a shorthand notation for a $(d+1)$-dimensional integral.
		It should be kept in mind in the following that $d$ only denotes the spatial dimensions, so the system is effectively $d$+1 dimensional.
		The bare propagators of the $h$ and $\overline{h}$ fields --- we do not call it a height field because of the many types of applications
		of the KS equation discussed above --- then write in Fourier space:
		\begin{equation}
		\label{eqG}
			\begin{split}
			&G^{h,\overline{h}}_0(\omega,\mathbf{q}) = \left<h(\omega,\mathbf{q})\overline{h}(-\omega,-\mathbf{q})\right> =
			\frac{1}{i\omega - \nu \mathbf{q}^2 + \zeta \mathbf{q}^4} \\
			&G^{h,h}_0(\omega,\mathbf{q}) = \left<h(\omega,\mathbf{q})h(-\omega,-\mathbf{q})\right> =
			\frac{2D(q)}{\big|i\omega - \nu \mathbf{q}^2 + \zeta \mathbf{q}^4\big|^2}
			\end{split}
		\end{equation}

	\subsection{A note on the role of the stochastic forcing}

		It appears in the reasoning above that the construction of the MSRJD action requires the introduction of a stochastic forcing in the KS equation.
		This is not totally satisfactory insofar as one of the most discussed properties of this system of equations is the property of spontaneous stochasticity,
		that was put in evidence in the founding paper of Kuramoto \cite{Kuramoto78}.
		By introducing the random force into the problem, we make the statistical properties of the solutions a priori dependent on the properties of the stochastic noise,
		and therefore loose one important piece of information.

		However, we would like to point out that in the original problem of Sivashinsky, the stochastic forcing was very much part of the problem under study.
		Also, both in the problem of the flow of a liquid film, or of chemical mixing, the inescapable presence of local inhomogeneities in the system do create
		a random forcing in all these problems in practise.
		Hence, as far as physics is concerned, it appears that the more fundamental problem is the random one.

		Besides, as discussed in the introduction, by a change of variable in the equation, typically subtracting the mean of the stochastic field, one
		gets from the equation without a right-hand side to a stochastically forced one.
		Generally, this procedure generates a noise that depends only on time, and not on space; but by generalising this method to local averages, the
		stochastic right-hand side depends on time and space.
		This proves that even the formulation of the KS problem without stochastic forcing is indeed equivalent to a forced one.
		The subtlety over which we went a bit too rapidly in this quick discussion is that such forcing is not totally random, it is produced from the
		solution to the equation, and may thus be related to it in a non trivial manner that is not captured by the usually studied class of noise terms;
		but if the chaotic solution decorrelates on sufficiently small length scales and time intervals, one can hope that the results derived in the following are
		not so different from those which would be obtained by studying a non forced version of the KS equation.

	\subsection{The Schwinger-Dyson formulation}

		In order to go further, we are going to need to make an assumption about the IR scaling of our solution.
		In agreement with previous studies on the KPZ, Burgers and related equations \cite{Gosteva24}, we parametrize the assumed stable IR regime
		with the help of two exponents defined as follows:
		\begin{equation}
			\nu(q)=\tilde{\nu}\,q^{-\eta_\nu}\ ,\ D(q) = \tilde{D}\,q^{-\eta_D}
		\end{equation}
		Furthermore, we are factoring out the $\eta_\nu$ dependent term, so that the renormalised propagator for the $h,\overline{h}$ channel for example
		can be rewritten as:
		\begin{equation}
		\label{eqGr}
			G^{h,\overline{h}}(\omega,\mathbf{q}) = \frac{q^{-\eta_\nu}}{i\tilde{\omega} - \tilde{\nu} q^2 + \tilde{\zeta} q^4}
		\end{equation}
		A similar equation holds for the renormalised propagator in the $h,h$ channel.
		Because there is no ambiguity, we will drop the $\tilde{\cdot}$ symbol in the following to lighten notations.

		Under these conditions, the Schwinger-Dyson equation writes:
		\begin{equation}
			G^{h,\overline{h}}(\Omega,\mathbf{p}) = G_0^{h,\overline{h}}(\Omega,\mathbf{p}) + G_0^{h,\overline{h}}(\Omega,\mathbf{p})\Pi(\Omega,\mathbf{p})
			G^{h,\overline{h}}(\Omega,\mathbf{p})
		\end{equation}
		where
		\begin{equation}
			\Pi(\Omega,\mathbf{p}) = \int\frac{d\omega}{2\pi}\int_{\mathbf{q}} \Gamma(\Omega,\mathbf{p},\omega-\Omega,\mathbf{q}-\mathbf{p})
			G^{h,h}(\omega,\mathbf{q}) G^{h,\overline{h}}(\Omega-\omega,\mathbf{p}-\mathbf{q})\Gamma(\Omega-\omega,\mathbf{p}-\mathbf{q},\omega,\mathbf{q})
		\end{equation}
		is the polarisation bubble, $\Gamma$ is the vertex function, and the shorthand notation for the wavenumber integral contains the appropriate number
		of $2\pi$ factors to comply with our Fourier convention.

		Regarding the vertex function $\Gamma$, we will require as little as we can so that our results stay general.
		Because the KS equation is translation invariant, we can write:
		\begin{equation}
			\Gamma(\Omega,\mathbf{p},\omega-\Omega,\mathbf{q}-\mathbf{p})\Gamma(\Omega-\omega,\mathbf{p}-\mathbf{q},\omega,\mathbf{q})
			 = \mathbf{q}^\alpha\mathbf{q}^\beta\gamma_{\alpha\beta}(q,p)
		\end{equation}
		This ensures that the bubble diagram always converges in the infrared (IR) limit.
		Furthermore, we require that $\gamma$ is an analytic function of its arguments, which means that the model stays well-defined across the scales,
		and that there is no emerging bound state at intermediate scales.

\section{The Edwards-Wilkinson regime}

	The Edwards-Wilkinson scaling regime corresponds to the particular scaling regime in which $\eta_\nu=\eta_D=0$.
	The short review in the introduction shows that this regime can be at least a weak attractor in the IR regime, hence, it is important that we study it.

	\subsection{General setup}

		If $\eta_\nu=0$, then $G=G_0$ for all types of upper indices, and the Schwinger-Dyson equation reduces to $\Pi(\Omega,\mathbf{p})=0$.
		Since $\eta_\nu=\eta_D=0$, the numerator does not get any more non trivial contributions.
		The structure of the denominator is given by (\ref{eqG}).
		The frequency integral can be evaluated first:
		\begin{equation}
		\label{eqEW1}
			\Pi(\Omega,\mathbf{p})=-\int_{\mathbf{q}}\frac{D(q)\mathbf{q}^\alpha \mathbf{q}^\beta\gamma_{\alpha\beta}(q,p)}
			{\mathbf{q}^2(\zeta \mathbf{q}^2 - \nu )(\zeta \mathbf{q}^4 -
			\nu \mathbf{q}^2 -\zeta (\mathbf{p}-\mathbf{q})^4 + \nu (\mathbf{p}-\mathbf{q})^2 -i\,\Omega)}
		\end{equation}
		Note that the IR limit is safe since the divergence of the denominator is compensated by the numerator.

		This integral is a bit challenging to compute because of the loss of the spherical symmetry.
		It is however possible to use, in $d>1$ (the generalisation to $d=1$ is obvious) the following decomposition of the vector $\mathbf{q}$:
		\begin{equation}
			\mathbf{q} = q_0 + \mathbf{q}_\perp\ ,\ q_0 = (\mathbf{q}\cdot\mathbf{p})\hat{\mathbf{p}}\ ,\ \mathbf{q}_\perp\cdot\mathbf{p}=0
		\end{equation}
		where the hatted vectors have a unit norm.
		With this decomposition, it is not difficult to show that (\ref{eqEW1}) is spherically symmetric in $\mathbf{q}_\perp$, a $(d-1)$-dimensional vector.

		As discussed above, there is no pole in 0 in this integral.
		Denoting by $S_d$ the surface of the $d$-dimensional sphere, after a suitable change of variable, the polarisation bubble writes:
		\begin{equation}
		\label{eqEW2}
		\begin{split}
			\Pi(\Omega,\mathbf{p})&=-\int_{q_0}\int_0^{+\infty}dy\,\frac{S_{d-1}y^{(d-1)/2-1}D(q_0,y)\Upsilon(q_0,y,p)}
			{(q_0^2+y)(\zeta (q_0^2 + y) - \nu )} \\
					      &\times\frac{1}{(\zeta (q_0^2 + y)^2 -
			\nu (q_0^2 + y) -\zeta (\mathbf{p}^2 -2 q_0\cdot \mathbf{p} + q_0^2 +y)^2 + \nu (\mathbf{p}^2 -2 q_0\cdot \mathbf{p} +q_0^2+ y) -i\,\Omega)}
		\end{split}
		\end{equation}
		where $\Upsilon$ is a shorthand notation for the $\gamma$ dependence that has to be separated into four terms depending on which components
		of the wave vector they are contracted to. The important point here is that this function, however complicated it may be, is not singular
		and therefore does not interfere with the pole structure of the integrand.
		It also has the property that the integral is UV safe, otherwise the Schwinger-Dyson equation itself would not be defined. Thus, on a contour
		at infinity, the integrand goes to zero sufficiently quickly.

		The $y$-integral can be computed by use of the residue theorem, remembering that $d=1$ has to be treated separately since in that case,
		$y$ is not defined (there is therefore no simple pole in 0).
		In the first line of (\ref{eqEW2}), two single poles can be easily identified: $y_1=-q_0^2$ and $y_2 = -q_0^2+ \nu/\zeta$.
		The denominator of the second line is more involved, but it boils down to a simple polynomial of order one in $y$, that therefore has only one root:
		\begin{equation}
			y_3 = \frac{\zeta\big( (\mathbf{p}^2 - 2q_0\cdot\mathbf{p})^2 -q_0^4\big) +\nu\big(q_0^2 - (\mathbf{p}^2- 2q_0\cdot\mathbf{p}) \big) + i\Omega}
			{2\zeta (q_0^2 - \mathbf{p}^2 + 2q_0\cdot\mathbf{p})}
		\end{equation}
		Note that thanks to the presence of the frequency term, this root never lies on the real axis.

\begin{figure}[h]
			\begin{center}
				\includegraphics[scale=1.4]{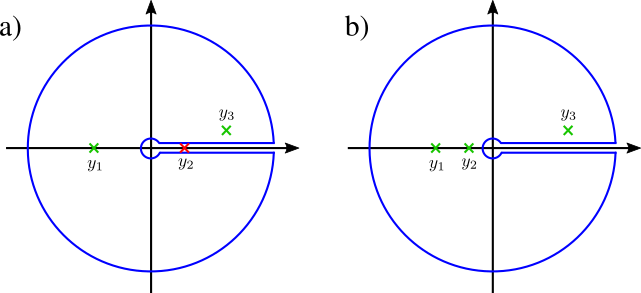}
			\end{center}
			\caption{a) Integration contour for $\nu>0$. There is one pole on the positive real axis.
			b) Integration contour for $\nu<0$. All the poles are inside the contour.}
			\label{figC4}
		\end{figure}

		The form of the integral allows to compute the $y$-integral through the use of the following contour: (i) a portion of circle of radius $\delta$ centered on
		the origin, that does not contribute in the limit $\delta\rightarrow0$; (ii) two straight lines at distance $\varepsilon$ of the real axis, that collapse onto
		the $y$-integral in the limit $\varepsilon\rightarrow0$; (iii) a portion of circle of radius $R$ that yields a negligible contribution in the limit
		$R\rightarrow +\infty$.
		This contour is represented on Fig.~\ref{figC4}.
		Note that the $y_3$ pole might be found in other places, for example in the lower half plane, depending on the value of the parameters, but what matters is that,
		whatever these values, $y_3$ always lies within the contour.

		Two scenarios are then to be distinguished: (i) $\nu>0$ (which corresponds to the initial KS problem): in this case, there always exists a value of $q_0$ such
		that $y_2$ lies on the positive part of the real axis (picture a) of Fig.~\ref{figC4}), so that the integrand is not integrable anymore, the $y$-integral
		diverges, and no IR stable solution exists; (ii) $\nu<0$ (which corresponds to the KPZ-like case): in this case, $y_2$ always lies on the negative
		part of the real axis and acts like a simple pole, the integral is well defined.
		The case $\nu=0$ is somewhat singular and is beyond the scope of the present study

		Hence, we have shown that, as far as the EW regime is concerned, there exists no stable IR scaling regime with $\nu>0$ (in the KS sense) --- this
		results can be extended easily to the case $d=1$ where, even though the splitting $q_0$/$y$ does not occur anymore, the divergence on the real axis
		appears in the same manner.

		We are, however, facing another challenge: the remaining $q_0$ integral can be evaluated through the residue theorem as well (the results are not shown
		since they are very lengthy and not necessary), and it is not difficult to convince oneself that, for generic values of the parameters of the problem,
		the $q_0$-integral does not vanish for $\nu<0$, which means that the polarisation bubble is not zero, and therefore, the EW equation leads to the paradox.
		Thus, there is no stable IR scaling regime for the EW scaling both for $\nu<0$ and for $\nu>0$.
		This result is compatible with the numerical study of Roy and Pandit \cite{Roy20}, though this does not constitute a formal proof.

	\subsection{The $\zeta\rightarrow0$ limit}

		The result of the previous section might seem satisfactory when analysed through the lens of the KS problem, because it is compatible with the numerical
		literature: as long as the system size is chosen to be large enough, the large scale statistics of the stationary state are of KPZ nature, and not EW.
		However, this raises a question: it has been shown \cite{Gosteva24} that the EW regime is an IR attractor of the dynamics of the Burgers-KPZ problem
		($\zeta=0$) when $d\geqslant2$.
		It is thus expected, if everything is consistent, that the polarisation bubble does vanish under these conditions, otherwise, if $\Pi$ is always non zero,
		that would mean that our previous result is a mere artefact of computation, and does not contain the appropriate physical content.

		Let us therefore examine the $\zeta\rightarrow0$ limit of the EW equations.
		In that case, the polarisation bubble becomes:

		\begin{figure}[h]
			\begin{center}
				\includegraphics[scale=1.4]{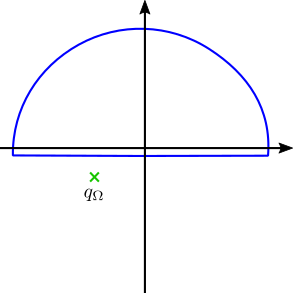}
			\end{center}
			\caption{Integration contour in the case where the pole lies on the lower half plane.
			If the pole is in the upper half plane, the contour is replaced by its image by a mirror symmetry around the real axis.}
			\label{figC3}
		\end{figure}

		\begin{equation}
		\label{eqEW3}
			\Pi(\Omega,\mathbf{p})=\int_{q_0}\int_0^{+\infty}dy\,\frac{S_{d-1}y^{(d-1)/2-1}D(q_0,y)\Upsilon(q_0,y,p)}
			{\nu\,(q_0^2+y)(\nu\, (\mathbf{p}^2 -2 q_0\cdot \mathbf{p}) -i\,\Omega)} 
		\end{equation}

		In the transverse integral, only the $y_1$ pole survives. It gives an a priori non-zero contribution.
		When evaluating the corresponding residue, the $(q_0^2 + y)$ term does not appear in the denominator anymore.
		Hence, the $q_0$ integral only contains one pole:
		\begin{equation}
			q_\Omega = \frac{i\Omega - \mathbf{p^2}}{2\sqrt{\mathbf{p^2}}}
		\end{equation}

		This pole has a well defined sign for its imaginary part (that depends on the value of the external parameters).
		As shown on Fig.~\ref{figC4}, it is thus always possible to build a contour such that the pole lies outside of it, namely, the $q_0$ integral in such conditions
		is equal to 0.
		We have therefore shown that when a transverse part for the momentum can be defined (that is $d\geqslant2$), the polarisation bubble is indeed equal to zero.
		In $d=1$, the polarisation integral is no longer defined in the UV limit and can therefore not be computed.
		We reach the well-known conclusion that the EW regime does not present a well defined IR scaling in $d=1$.

		Let us pause one moment to appreciate what has been shown so far.
		Under the only hypothesis of the existence of a scaling regime in the IR, we have shown that the EW regime, correctly identified with respect to the previous
		literature on the subject, becomes unstable to the presence of the $\partial^4h$ term in the infrared regime, irrespective of the sign of the viscosity
		parameter $\nu$.
		However, its role as an attractor in the IR for $d=2$ in the limit $\zeta\rightarrow0$ certainly explains why phase space trajectories might spend quite an
		amount of time in its vicinity, and thus why large length and time scales are required in practise to well resolve the KPZ-statistics regime in the
		KS equation
		(recall that our theoretical setup is by construction adapted to this kind of constraints, which is not so obvious when going to numerical implementation of the
		equation).

\section{The KS-KPZ regime}

	In this section, we investigate the full KS-KPZ regime, with non trivial critical exponents.
	Before beginning, let us remind our goal here: we search for general properties of the KS equation in the IR limit under a \textit{minimal} set of hypotheses;
	hence, even if it is possible to extract the values of the exponents from the Schwinger-Dyson equations, it is only possible to do so in an appropriate way
	by numerically solving a self-consistent equation.
	The quantitative evaluation of the exponents becomes then more a reflect of the precision of the numerical method used to derive them rather than of the
	Schwinger-Dyson setup itself.
	As a result, solving for the values of the critical exponents goes beyond the scope of the present study.
	We will simply suppose $\eta_\nu\neq0$, $\eta_D\neq0$.

	From (\ref{eqGr}), the IR domain is now dominated on the left-hand side by the renormalised propagator, and the right-hand side by the bubble term,
	the bare propagator being negligible in this limit.
	The expression of the bubble now writes:

	\begin{equation}
		\label{eqEW4}
		\begin{split}
			\Pi(\Omega,\mathbf{p})&=-\int_{q_0}\int_0^{+\infty}dy\,\frac{S_{d-1}y^{(d-1)/2-1}D(q_0,y)\Upsilon(q_0,y,p)}
			{(q_0^2+y)(\zeta (q_0^2 + y) - \nu )} \\
					      &\times\frac{(q_0^2+y)^{-\eta_D/2+3\eta_\nu/2}}{(\zeta (q_0^2 + y)^2 -
			\nu (q_0^2 + y) -\zeta (\mathbf{p}^2 -2 q_0\cdot \mathbf{p} + q_0^2 +y)^2 + \nu (\mathbf{p}^2 -2 q_0\cdot \mathbf{p} +q_0^2+ y) -i\,\Omega)}
		\end{split}
	\end{equation}

	The well definiteness of the integral in the IR leads to the inequality:
	\begin{equation}
		d- 4-\eta_D + 3\eta_\nu > -1
	\end{equation}
	which interestingly is saturated precisely at the KPZ fixed point \cite{Gosteva24}.

	We then proceed in the same way as for the EW regime: we evaluate the $y$-integral using the contour on Fig.~\ref{figC3}, and two cases are possible, either 
	$\nu>0$ and one pole lies on the real axis, in which case the integral diverges and there are no IR stable regime possible, or $\nu<0$ and the three poles lie
	within the contour, and the polarisation bubble stays finite.
	The evaluation of the $q_0$-integral can be made in the same way with a contour similar to that of Fig.~\ref{figC4}, except that now the structure of the poles
	is much more involved.

	As explained above, providing a numerical solution for the values of $\eta_\nu$ and $\eta_D$ as a function of $d$ does not fit the scope of our study.
	What we have shown, on the other hand, is that the only possibility to get an IR stable regime is to have $\nu<0$, that is to get KPZ statistics.
	This result is in agreement with most of the previous literature on the KS equation which focused on the $d=1$ case.
	Here, we have shown that, contrary to the claim of Yakhot, the change of sign of $\nu$ upon renormalisation in the KS equation is a generic property of this
	equation that holds in any dimension (or at least the physically relevant ones), and is not restricted to $d=1$.

\section{Conclusions}

	All in all, we have proven in this study that under the sole assumption that a IR stable scaling regime exists, it can only belong to the KPZ universality class,
	whatever the initial conditions --- except for the inviscid case $\nu=0$ that we keep for a later study --- provided the time and length scales are large enough.
	This result is in agreement with the numerical literature in $d=1$, but contradicts the prediction of Yakhot that such effect only occurs in $d=1$.
	We hope that our result will help motivate further studies on the studies in $d>1$ where, since the prediction of Yakhot results have been very scarce.

\section*{Acknowledgements}

	I am grateful to  E. Simonnet, C. Campolina, L. Long and M. Massaro for enlightening discussions.
	I would also like to thank L. Canet and L. Gosteva for kindly sharing their knowledge on this problem.

\bibliography{EcGSNSI.bib}

\end{document}